\documentclass[aps,prd,amsmath,amssymb,showpacs]{revtex4}
\usepackage{graphicx}
\begin{document}

\title{Extra dimensions in photon-induced two lepton
final states at the CERN-LHC}

\author{S. Ata\u{g} }
\email[]{atag@science.ankara.edu.tr}
\affiliation{Department of Physics, Faculty of Sciences,
Ankara University, 06100 Tandogan, Ankara, Turkey}

\author{S.C. \.{I}nan}
\email[]{sceminan@cumhuriyet.edu.tr}
\affiliation{Department of Physics, Cumhuriyet University,
58140, Sivas, Turkey}

\author{\.{I}. \c{S}ahin}
\email[]{inancsahin@karaelmas.edu.tr} \affiliation{Department of
Physics, Zonguldak Karaelmas University, 67100 Zonguldak, Turkey}

\begin{abstract}
We discuss the potential of the photon-induced two lepton
final states at the LHC to explore the phenomenology of the
Kaluza-Klein (KK) tower of gravitons in the scenarios of the
Arkani-Hamed, Dimopoulos and Dvali(ADD) model
and Randall-Sundrum (RS) model. The sensitivity to model
parameters can be improved compared to the present LEP or
Tevatron sensitivity.
\end{abstract}


\maketitle

\section{Introduction}
The general purpose Large Hadron Collider (LHC) experiments
ATLAS and CMS at CERN   have central detectors with a pseudorapidity
$\eta$ coverage 2.5 for tracking system and 5.0 for calorimetry.
However, the significant amount of particles and  energy flow
are in the very forward directions and  not detected by these
detectors. Moreover, some parts of the cross sections measured
by the central detectors may belong to the elastic scattering
and ultraperipheral collisions. For deeper understanding of the
physics from very forward region   additional equipments are needed.
ATLAS and CMS collaborations have a program of forward physics
with extra detectors located in a region nearly 100m-400m from the
interaction point \cite{royon}. These forward detectors will be installed
as close as a few mm to the beamline. The physics program of this
new instrumentation covers soft and hard diffraction,
high energy photon induced interactions,
low-x dynamics with forward jet studies,
large rapidity gaps between forward jets,
and luminosity monitoring \cite{royon,khoze, schul}.
One of the main goal of these forward
detectors is to tag the protons  with some energy fraction loss
$\xi =E_{loss}/E_{beam}$.
Specifically, ATLAS and CMS in standard running conditions will have
forward detectors positioned at 220m and 420m from interaction point
with an acceptance of $0.0015<\xi<0.15$ \cite{royon2,albrow}.
TOTEM experiment at 147m
and 220m from CMS interaction point together with
standard CMS-TOTEM at 420m  have an overall acceptance
$0.0015<\xi<0.5$. When the  forward detectors are installed
closer to the interaction points, higher $\xi$ is obtained.

One of the well known applications of the forward detectors
is the high energy photon induced interaction with exclusive
two lepton ($e^{+}e^{-}$ or $\mu^{+}\mu^{-}$) final states.
Two quasi-real photons emitted by each proton interact
each other to produce two leptons
$\gamma\gamma \to \ell^{+}\ell^{-}$.
Deflected protons and their energy loss will be detected
by the forward detectors mentioned above
but leptons will go to central detector. Charged leptons
with rapidity $|\eta|<2.5$ and $p_{T}>(5-10) GeV$ will be identified
by the central detector. Photons emitted with small angles
by the protons show  a spectrum of virtuality $Q^{2}$ and energy
$E_{\gamma}$. This is described by the equivalent photon
approximation \cite{budnev,baur} which differs from the
point-like electron positron case by taking care of
the electromagnetic form factors
in the equivalent photon spectrum and  effective
$\gamma\gamma$ luminosity

\begin{eqnarray}
dN=\frac{\alpha}{\pi}\frac{dE_{\gamma}}{E_{\gamma}}
\frac{dQ^{2}}{Q^{2}}[(1-\frac{E_{\gamma}}{E})
(1-\frac{Q^{2}_{min}}{Q^{2}})F_{E}+\frac{E^{2}_{\gamma}}{2E^{2}}F_{M}]
\end{eqnarray}

where

\begin{eqnarray}
Q^{2}_{min}=\frac{m^{2}_{p}E^{2}_{\gamma}}{E(E-E_{\gamma})},
\;\;\;\; F_{E}=\frac{4m^{2}_{p}G^{2}_{E}+Q^{2}G^{2}_{M}}
{4m^{2}_{p}+Q^{2}} \\
G^{2}_{E}=\frac{G^{2}_{M}}{\mu^{2}_{p}}=(1+\frac{Q^{2}}{Q^{2}_{0}})^{-4},
\;\;\; F_{M}=G^{2}_{M}, \;\;\; Q^{2}_{0}=0.71 \mbox{GeV}^{2}
\end{eqnarray}

Here E is the energy of the proton beam which is related to the
photon energy by $E_{\gamma}=\xi E$
and $m_{p}$ is the mass of the proton.
The magnetic moment of the proton is $\mu^{2}_{p}=7.78$, $F_{E}$ and
$F_{M}$ are functions of the electric and magnetic form factors.
The cross section $d\sigma_{\gamma\gamma \to \ell\ell}$
for the subprocess $\gamma\gamma \to \ell^{+}\ell^{-}$
should be integrated over the photon spectrum

\begin{eqnarray}
d\sigma=\int{\frac{dL^{\gamma\gamma}}{dW}}
d\sigma_{\gamma\gamma \to \ell\ell}(W)dW
\end{eqnarray}

where the effective photon luminosity $dL^{\gamma\gamma}/dW$
is given by

\begin{eqnarray}
\frac{dL^{\gamma\gamma}}{dW}=\int_{Q^{2}_{1,min}}^{Q^{2}_{max}}
{dQ^{2}_{1}}\int_{Q^{2}_{2,min}}^{Q^{2}_{max}}{dQ^{2}_{2}}
\int_{y_{min}}^{y_{max}}
{dy \frac{W}{2y} f_{1}(\frac{W^{2}}{4y}, Q^{2}_{1})
f_{2}(y,Q^{2}_{2})}.
\end{eqnarray}
with

\begin{eqnarray}
y_{min}=\mbox{MAX}(W^{2}/(4\xi_{max}E), \xi_{min}E), \;\;\;
y_{max}=\xi_{max}E, \;\;\;
f=\frac{dN}{dE_{\gamma}dQ^{2}}.
\end{eqnarray}
Here W is the invariant mass of the two photon system
$W=2E\sqrt{\xi_{1}\xi_{2}}$ and maximum virtuality is
$Q^{2}_{max}=2$ $\mbox{GeV}^{2}$.

The lepton pair production by two photon fusion yields very clean final
states and it is convenient to use it as a luminosity monitor for the LHC.
As it  was discussed in Ref.\cite{khoze2}
the main possible contamination comes
from the proton dissociation into X, Y system,
$pp \to X+\ell^{+}\ell^{-}+Y $ where X and Y are baryon excitations
such as $N^{*}$, $\Delta$ isobars.
To reduce this contribution it was proposed to impose a cut on the
transverse momentum of the photon pair
$|\vec{q}_{1t}+\vec{q}_{2t}|<(10-30)$MeV. In
actual experiment this cut should be applied on the lepton
pair. It is shown in some detail in ref.\cite{khoze2}
that the cut of 30 MeV on the
transverse momentum of the lepton pair is possible although a cut
of at least 5 GeV is needed for individual lepton detection.
However, in theoretical calculations,
if one uses equivalent photon approximation the
subprocess $\gamma\gamma \to \ell^{+}\ell^{-}$ is factorized.
In this case, this cut can be applied through effective
photon luminosity using the relation between $Q^{2}$ and transverse
momentum of the photon $q_{t}$

\begin{eqnarray}
Q^{2}=Q^{2}_{min}+\frac{q^{2}_{t}}{1-E_{\gamma}/E}
\end{eqnarray}

\begin{figure}
\includegraphics{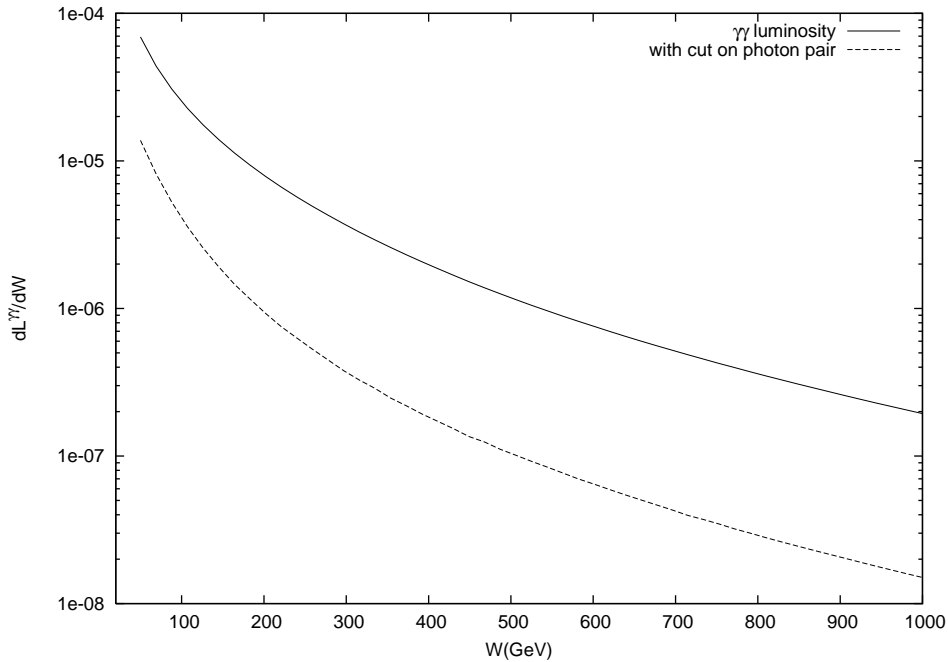}
\caption{Effective $\gamma\gamma$ luminosity as a function
of the invariant mass of the two photon system. Lower
curve represents the case with a cut on the total momentum
of the photon pair
$|\vec{q}_{1t}+\vec{q}_{2t}|<30$MeV.
\label{fig1}}
\end{figure}

This is why we have kept integration over
$Q^{2}$ in the luminosity expression above.
In Fig.\ref{fig1} effective $\gamma\gamma$ luminosity
is shown with and without transverse momentum cut imposed
on the photon pair.

It is clear that the photon-induced two lepton final states with
invariant di-photon mass $W > 1$ TeV seems highly attractive to
probe new physics beyond the Standard Model(SM) with available
luminosity.
In this work, we explore the phenomenology of extra dimensions
in the framework of  the Arkani-Hamed,
Dimopoulos and Dvali(ADD) model of
large extra dimensions and Randall-Sundrum (RS) model
of warped extra dimensions via the photon induced process
$pp \to p\ell^{+}\ell^{-} p$ at the LHC.

\section{ADD Model of Large Extra Dimensions}

It is well known that there is a large  difference
in energy between the electroweak scale
which is a few hundreds of GeV and gravity
scale that is the Planck scale $M_{Pl}\sim 10^{19}$GeV
in a four dimensional spacetime.
This is called the hierarchy problem in particle
physics. In string theory, extra space dimensions
higher than three has  already been contained.
Following the string theory ideas,
three space dimensional world is called a "wall" or
3-brane where all Standard Model particles are
confined to this wall. The D-dimensional spacetime,
$D=3+\delta+1$, with $\delta$ extra space dimensions
is called "bulk" where 3-brane is embedded in it.
The way how to handle hierarchy problem depends on the
models.

According to the model proposed by
Arkani-Hamed, Dimopoulos and Dvali   SM fields
can not go out of the 3-brane while gravity propagates
in the bulk \cite{add}. In D dimensions, the solutions of the
linearized equations of motion of the metric field are
the Kaluza-Klein tower. After integrations over extra
dimensions, resulting 4-dimensional fields are the
Kaluza-Klein modes. KK zero mode field is massless
corresponding to the graviton in 4-dimensions but KK excited
modes are massive. In this model extra dimensions are
compactified with a compactification radius
$r_{c}\sim $mm-fermi (or $1/r_{c}\sim $eV-MeV)
which determines the KK mode spacing.
This mode spacing is very small when compared to
typical collider energies which allows the summation
over large number of KK states in the final states
or in the propagator. Because of this large compactification
radius, ADD model is
known as the large extra dimension model.
The overall effect of the KK states makes the gravity
strong in $D=4+\delta$ dimensions and thus its effective
interactions in 4-dimensions  with the
Standard Model particles  are sizable at
collider energies. It is possible to relate the Planck Mass
$M_{Pl}$ to the corresponding scale $M_{D}$ in D-dimensions
through the compactified volume $V_{\delta}$ by the relation

\begin{eqnarray}
M^{2}_{Pl}=V_{\delta}M^{2+\delta}_{D}
\end{eqnarray}

It is assumed that  $M_{D}$ is in TeV
region and $M_{Pl}$ becomes
large due to large higher dimensional  volume
of $V_{\delta}$ with $\delta=2-7$. This situation suggests
that Planck scale $M_{Pl}$ is no longer fundamental scale.
However, the large gap between the electroweak
and Planck scale is compensated by the large
compactification scale of the extra dimensions.

Let us now calculate the Feynman amplitude for the subprocess
$\gamma\gamma \to \ell^{+}\ell^{-}$ by adding an
s-channel diagram  with KK graviton exchange to the
t and u channel diagrams of electromagnetic interaction \cite{add2}.

\begin{eqnarray}
iM_{KK}=\sum_{n}[\bar{u}(p_{1})\Gamma_{2}^{\mu\nu} v(p_{2})
\frac{\frac{i}{2}B_{\mu\nu\alpha\beta}}{\hat{s}-m^{2}_{n}}
\Gamma_{1}^{\alpha\beta\rho\sigma} e_{\rho}(k_{1})e_{\sigma}(k_{2})]
\end{eqnarray}
where $k_{1}, k_{2}$, $p_{1}, p_{2}$ and $e_{\rho}(k_{i})$
are incoming photon, outgoing lepton momenta and
polarization vectors of photons. The coherent sum is over KK modes.
Vertex functions
$\Gamma_{1}^{\alpha\beta\rho\sigma}$ for $KK-\gamma\gamma$
and $\Gamma_{2}^{\mu\nu}$ for $KK-\ell\ell$ are given below

\begin{eqnarray}
\Gamma_{1}^{\alpha\beta\rho\sigma}=-\frac{i\kappa}{2}
[(k_{1}\cdot k_{2})C^{\alpha\beta\rho\sigma}+
D^{\alpha\beta\rho\sigma}]
\end{eqnarray}

\begin{eqnarray}
\Gamma_{2}^{\mu\nu}=-\frac{i\kappa}{8}
[\gamma^{\mu}(p^{\nu}_{1}-p^{\nu}_{2})+
\gamma^{\nu}(p^{\mu}_{1}-p^{\mu}_{2})]
\end{eqnarray}
coupling constant $\kappa$ is related to the Newton constant
$G^{4+\delta}_{N}$ in $D=4+\delta$ dimension by
$\kappa^{2}=16\pi G^{4+\delta}_{N}$.
Explicit forms of the tensors $C^{\alpha\beta\rho\sigma}$ and
$D^{\alpha\beta\rho\sigma}$ are given in the appendix.
Tensor $B_{\mu\nu\alpha\beta}$ in the propagator of KK graviton
is

\begin{eqnarray}
B_{\mu\nu\alpha\beta}=\eta_{\mu\alpha}\eta_{\nu\beta}+
\eta_{\mu\beta}\eta_{\nu\alpha}-\frac{2}{3}
\eta_{\mu\nu}\eta_{\alpha\beta}
\end{eqnarray}
where $\eta_{\mu\nu}$ is the metric tensor of the flat space
in 4-dimensions.

The summation over Kaluza-Klein modes can be calculated without
specifying any specific process. Since the KK tower is an infinite
sum, ultraviolet divergences are present in tree level process. Thus
we need a cutoff procedure.
For phenomenological applications following result was obtained
in the literature \cite{add2}

\begin{eqnarray}
\kappa^{2}D(\hat{s})&&\equiv
\kappa^{2}\sum_{n}\frac{i}{\hat{s}-m^{2}_{n}}=
\frac{-16\pi}{\Lambda^{4}_{c}} \;\;\;\;\;\mbox{for} \;\; \delta > 2 \\
\end{eqnarray}
where  $\Lambda_{c}$ is the cutoff energy. The expression connecting
the unknown cutoff energy $\Lambda_{c}$ to the fundamental scale
$M_{D}$ is not known without the knowledge of full theory.
The relation $\Lambda_{c} < M_{D}$ can be written based on the
string theory. In practical calculation the equality
$M_{D}\simeq \Lambda_{c}$ corresponds to the lower limit
of the fundamental scale $M_{D}$. In every part of this work
containing the KK graviton propagator of ADD model
we set $M_{D}\simeq \Lambda_{c}$ and limits
on $M_{D}$ always mean its lower limit.

\begin{figure}
\includegraphics{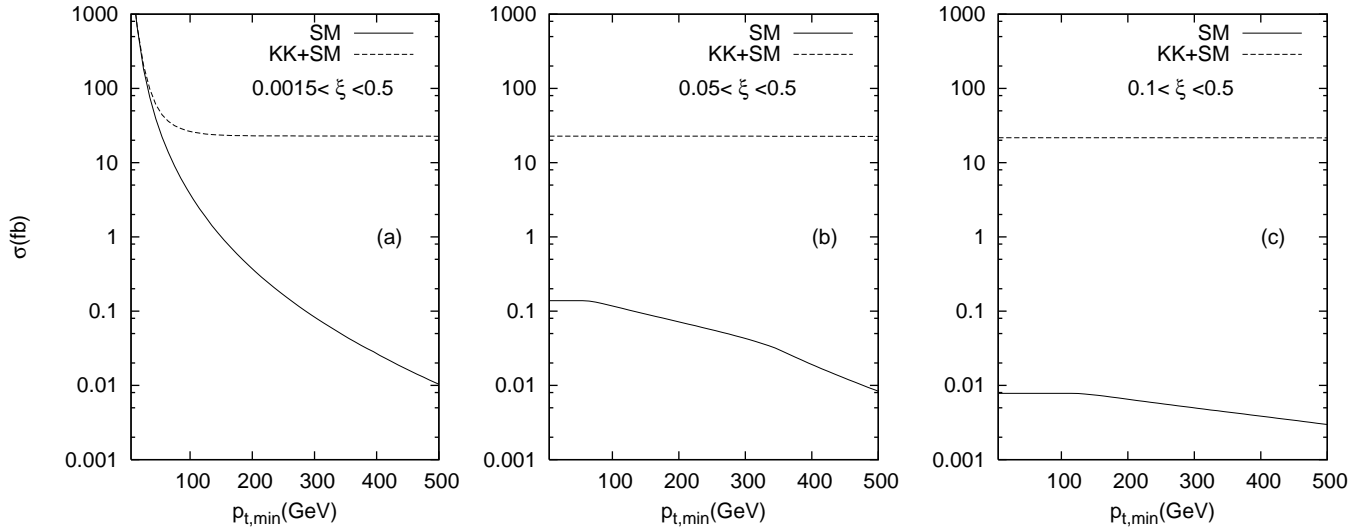}
\caption{Cross section of the process $pp \to p\ell^{+}\ell^{-}p$
as a function of the transverse momentum cut of the final leptons
with and without KK graviton exchange for $M_{D}=1500$ GeV and for
three acceptance regions of forward detectors.
\label{fig2}}
\end{figure}

\begin{figure}
\includegraphics{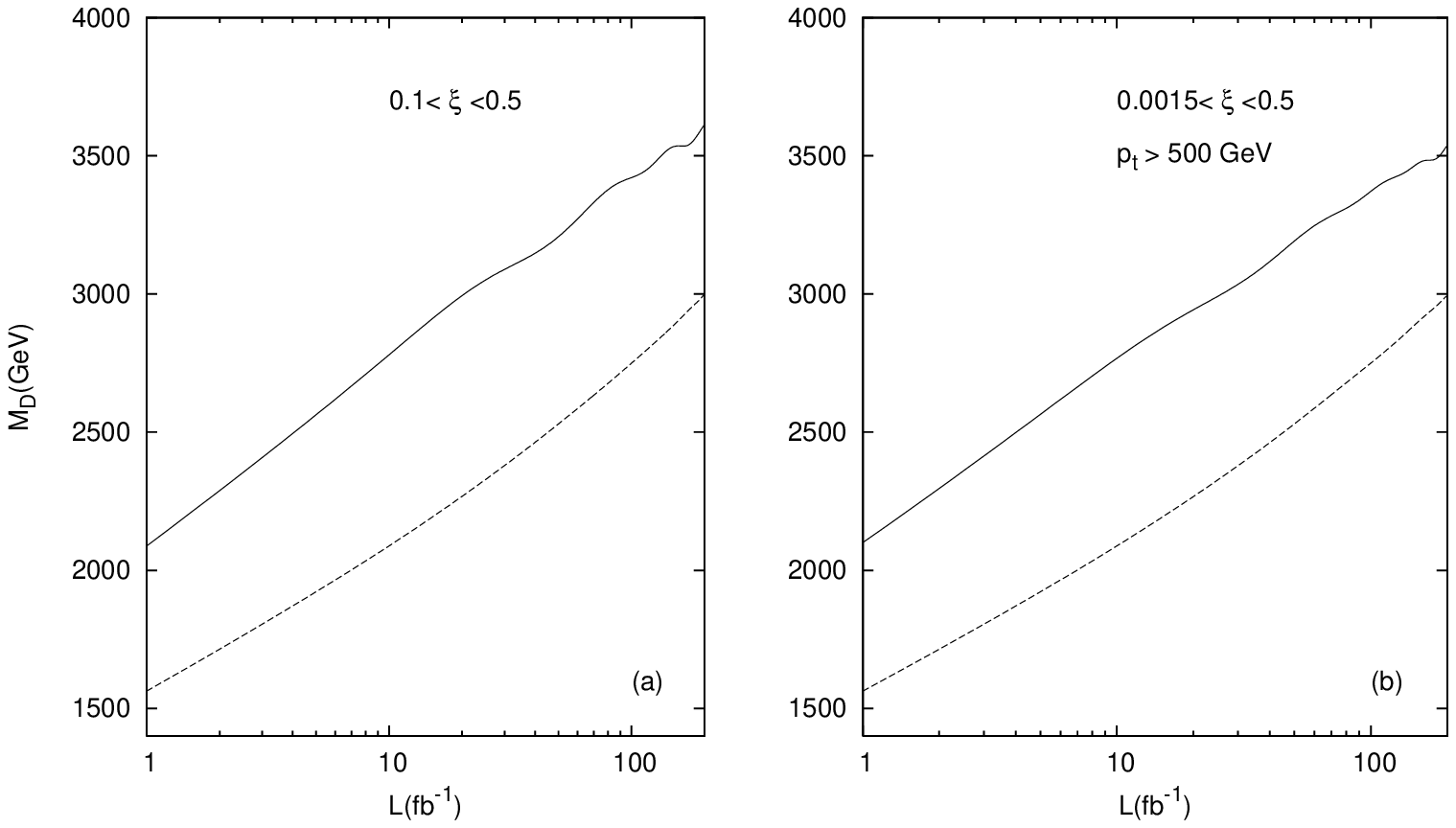}
\caption{95\% C.L. search reach for $M_{D}$ as a function of
integrated LHC luminosity for two acceptance regions shown
in plots (a) and (b). In the  plot (b), a cut of 500 GeV is imposed
on the transverse momenta of the  final state leptons.
Lower curves are the results from the cut on the transverse
momentum of the incoming photon pair.
\label{fig3}}
\end{figure}

The whole squared amplitude which consists of electromagnetic,
KK and interference parts has been calculated in terms of
Mandelstam invariants $\hat{s}$ and $\hat{t}$, neglecting lepton
masses

\begin{eqnarray}
|M|^{2}&&=|M_{em}|^{2}+|M_{KK}|^{2}+|M_{int}|^{2}, \\
|M_{em}|^{2}&&=-8g^{4}_{e}
[\frac{\hat{s}+\hat{t}}{\hat{t}}+
\frac{\hat{t}}{\hat{s}+\hat{t}}], \\
|M_{KK}|^{2}&&=|\kappa^{2}D(\hat{s})|^{2}[-\frac{\hat{t}}{8}
(\hat{s}^{3}+2\hat{t}^{3}+3\hat{t}\hat{s}^{2}+
4\hat{t}^{2}\hat{s})], \\
|M_{int}|^{2}&&=-2g^{2}_{e}\kappa^{2}\frac{1}{2}
(D(\hat{s})+D^{\star}(\hat{s}))
[\hat{s}^{2}+2\hat{t}^{2}+2\hat{s}\hat{t}],
\;\;\;\;\; g^{2}_{e}=4\pi\alpha
\end{eqnarray}
where the factor due to initial spin average is absent.
Starting from Fig.\ref{fig3} to the rest of the work
the cross sections will be multiplied by a factor of 2 to
include both electrons and muons in the final states.
Collider signals of virtual graviton exchange may lead to the
deviations from the SM in the magnitude of the cross section
and in the angular distributions of the final particles
due to spin-2 nature of the graviton. First we show the
the contribution
of the KK graviton to the total cross section of the main process
$pp \to p \ell^{+}\ell^{-}p$ in the photon induced interactions
at LHC with forward detectors. We consider three acceptance
regions of the  forward detectors
$0.0015<\xi<0.5$,\;\; $0.05<\xi<0.5$ and $0.1<\xi<0.5$ to see the
comparison. Fig.\ref{fig2}  shows
the cross sections with and without KK graviton exchange versus
the minimum transverse momenta (or $p_{t}$ cut) of the final leptons
for $M_{D}=1500$ GeV. It is clear from the Fig.\ref{fig2} that the
deviation from the SM gets higher as the $p_{t}$ cut increases.
An important thing that has to be stressed here  is the correlation
between the acceptance region and the $p_{t}$ cut. When Fig.\ref{fig2}(a)
and Fig.\ref{fig2}(c) are compared, we see that the acceptance region
$0.1<\xi<0.5$ has almost the same feature as the region
$0.0015<\xi<0.5$ with $p_{t,min}\sim 500$ GeV. This means that
a high lower bound of the acceptance region mimics an extra
$p_{t}$ cut.

To see  this explicitly, we calculated 95\% C.L.  bounds
on the $M_{D}$ as a function of the integrated LHC luminosity
for  two acceptance regions $0.1<\xi<0.5$ with
$p_{t}>10$ GeV and $0.0015<\xi<0.5$ with $p_{t}> 500$ GeV. In
both cases we kept $|\eta|<2.5 $.
Because of the small number of SM events in these regions
(due to high $p_{t}$ cut in the second region)
Poisson distributed events are considered for statistical
analysis. In confidence level analysis with Poisson distributed
data,  number of observed events is assumed to be equal
to the SM prediction $N_{obs}=\sigma^{SM}L$ for an integrated
luminosity $L$. The estimations are shown in Fig.\ref{fig3}(a)
and Fig.\ref{fig3}(b) which have nearly the same behaviour
as each other.
Therefore, taking acceptance region $0.1<\xi<0.5$ is
good enough to feel the KK graviton exchange in the process
$pp \to p\ell^{+}\ell^{-} p$ via two photon fusion.
When wider acceptance regions with smaller lower bound
is considered it is far better to use high lepton $p_{t}$ cut
to supress SM contributions.

\begin{figure}
\includegraphics{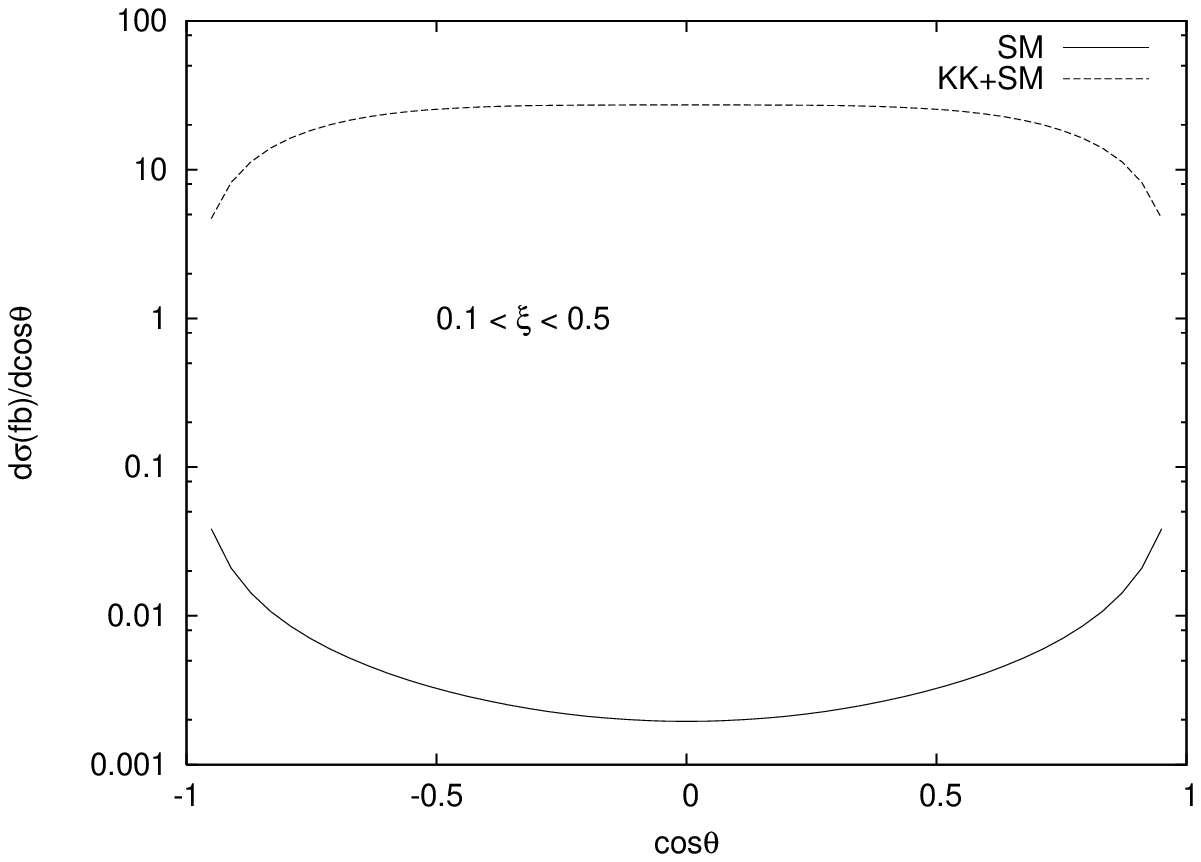}
\caption{Angular distribution of the final leptons
for the acceptance region $0.1 < \xi < 0.5$
and $M_{D}=1500$ GeV. The contribution of KK gravitons
is clearly separated from the SM part in magnitude
and in shape.
\label{fig4}}
\end{figure}

Angular behaviour of the final leptons
is given in Fig.\ref{fig4}
in the center of mass frame of the the subprocess
for the acceptance region $0.1 < \xi < 0.5$
and $M_{D}=1500$ GeV. The difference between
the KK graviton contribution and the SM part is sizable
in magnitude and in shape.

\section{RS Model of Warped Extra Dimensions}

An alternative way to handle the large  hierarchy between electroweak
and gravity scales is the one proposed by Randall and Sundrum
in which curvature in higher dimensions is responsible for removing
this large hierarchy. The metric in 5D is obtained as a solution
to Einstein's equations under the 4D Poincare invariance \cite{rs}

\begin{eqnarray}
ds^{2}=e^{-2ky}\eta_{\mu\nu}dx^{\mu}dx^{\nu}-dy^{2},
\end{eqnarray}
where y is the $5^{th}$ dimension which is parametrized as
$y=r_{c}|\Phi|$ with $r_{c}$ being the compactification radius
of the extra dimension. Angular coordinate $\Phi$
has the limits of $0\leq|\Phi|\leq\pi$.
First term in this metric is the 4D Minkowski spacetime
metric  multiplied by an exponential factor that
is called warp factor containing fifth
dimension and the degree of the curvature k.
In this model, two 3-brane with
opposite and equal tensions stand at the boundaries
of  a 5D  Anti-de-Sitter space.
Each boundary includes 4D Minkowski
metric and y is orthogonal to each 3-brane.
The distance between the two walls is $y=\pi r_{c}$.
Gravity propagates in $5^{th}$ dimension and
localized on the wall at $y=0$ called Planck brane.
The other wall at $y=\pi r_{c}$ is referred to
TeV brane where SM fields live on.
Starting with 5D action, interaction of the KK gravitons
with the matter fields can be found in the 4D effective
lagrangian

\begin{eqnarray}
L=-\frac{1}{\bar{M}_{Pl}}T^{\alpha\beta}(x)h^{(0)}_{\alpha\beta}(x)
-\frac{1}{\Lambda_{\pi}}T^{\alpha\beta}(x)
\sum_{n=1}^{\infty}h^{(n)}_{\alpha\beta}(x)
\end{eqnarray}
where $T^{\alpha\beta}(x)$ is the energy momentum tensor of the
matter field in the Minkowski space and
$\bar{M}_{Pl}=M_{Pl}/\sqrt{8\pi}$ is the reduced Planck scale.
$h^{(n)}_{\alpha\beta}(x)$ indicates the KK modes of
the graviton on the 3-brane. It is seen that the masless
zero KK mode decouples from the sum and its coupling
strength is $1/\bar{M}_{Pl}$. The massive  KK states has the
coupling of $(1/\Lambda_{\pi})\sim 1/\mbox{TeV}$ with
$\Lambda_{\pi}=e^{-kr_{c}\pi}\bar{M}_{Pl}$. The
4D effective action based on the RS model leads to the relation
between 5D fundamental Planck scale M and usual 4D reduced
Planck scale

\begin{eqnarray}
\bar{M}^{2}_{Pl}=\frac{M^{3}}{k}(1-e^{-2kr_{c}\pi}).
\end{eqnarray}
It is assumed that $k\sim M_{Pl}$ then the above relation
states $M\sim M_{Pl}$. Therefore, there is no
additional hierarchy created by the model. When
$kr_{c}$ is taken to be  $\sim 10-12$  all the physical
processes happen in TeV scale on TeV-brane. This fact
states that  the hierarchy is generated by the warp factor.

\begin{figure}
\includegraphics{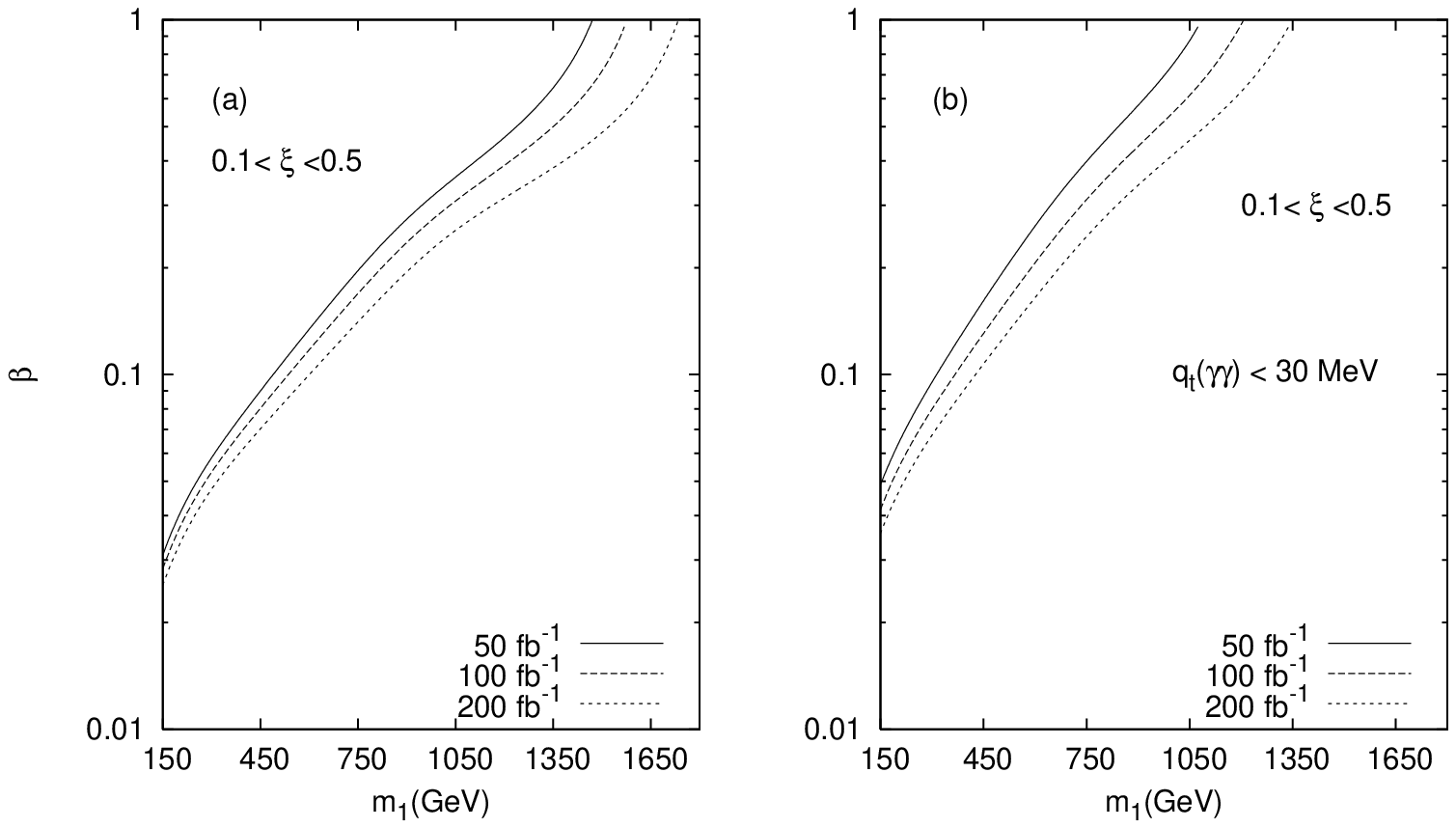}
\caption{95\% C.L. exclusion region for the parameters
$\beta$ and $m_{1}$ at three different integrated LHC
luminosities; $50 fb^{-1}$, $100 fb^{-1}$ and $200 fb^{-1}$.
Curves in the  Fig.(b)  are obtained with
the cut on the transverse momentum of the
incoming photon pair. Excluded regions are defined
by the area over the curves.
\label{fig5}}
\end{figure}

The mass spectrum of the  KK modes of the graviton
in the RS model  is given by \cite{rs}

\begin{eqnarray}
m_{n}=x_{n}ke^{-kr_{c}\pi}=x_{n}\beta\Lambda_{\pi},
 \;\; \mbox{with}\;\; \beta=\frac{k}{\bar{M}_{Pl}}
\end{eqnarray}
where $x_{n}$ are the roots of the Bessel function of order 1
$J_{1}(x_{n})=0$. The first values are $x_{1}=3.83$, $x_{2}=7.02$
and $x_{3}=10.17$. The scale of the  masses is
$m_{n}\sim \Lambda_{\pi}\sim \mbox{TeV}$
on the ground of   $\beta\sim 1$. This fact demonstrates that
the effect of each excitation should be seen separately
at colliders. The RS scenario is described by two parameters
$\Lambda_{\pi}$ and $\beta$ which are appropriate for phenomenological
applications.
In the squared amplitude for $\gamma\gamma \to \ell^{+}\ell^{-}$
the only difference from the ADD case
occurs in the following form of the graviton propagator

\begin{eqnarray}
\kappa^{2}D(\hat{s})=\frac{2}{\Lambda^{2}_{\pi}}
\sum_{n=1}^{\infty}\frac{1}{\hat{s}-m^{2}_{n}+i\Gamma_{n}m_{n}},
\;\;\;\;\;  \Gamma_{n}=\rho m_{n}(\frac{m_{n}}{\Lambda_{\pi}})^{2}
\end{eqnarray}
with $\rho=1$ is used in the width $\Gamma_{n}$ of the individual
KK graviton.
Constraint on the parameter $\beta$ can be found for the first graviton
mode  with mass $m_{1}$. The estimation for 95\% C.L. parameter
exclusion region is shown in Fig.\ref{fig5} for the integrated LHC
luminosities; $50 fb^{-1}$, $100 fb^{-1}$ and $200 fb^{-1}$,
with the acceptance region $0.1 <\xi < 0.5$.

In the approximation $m^{2}_{n}>>s$ \;\; KK graviton propagator takes
the form

\begin{eqnarray}
\kappa^{2}D(\hat{s})=\frac{2}{\beta^{2}\Lambda^{4}_{\pi}}
\sum_{n=1}^{\infty}\frac{-1}{x^{2}_{n}}.
\end{eqnarray}

\begin{figure}
\includegraphics{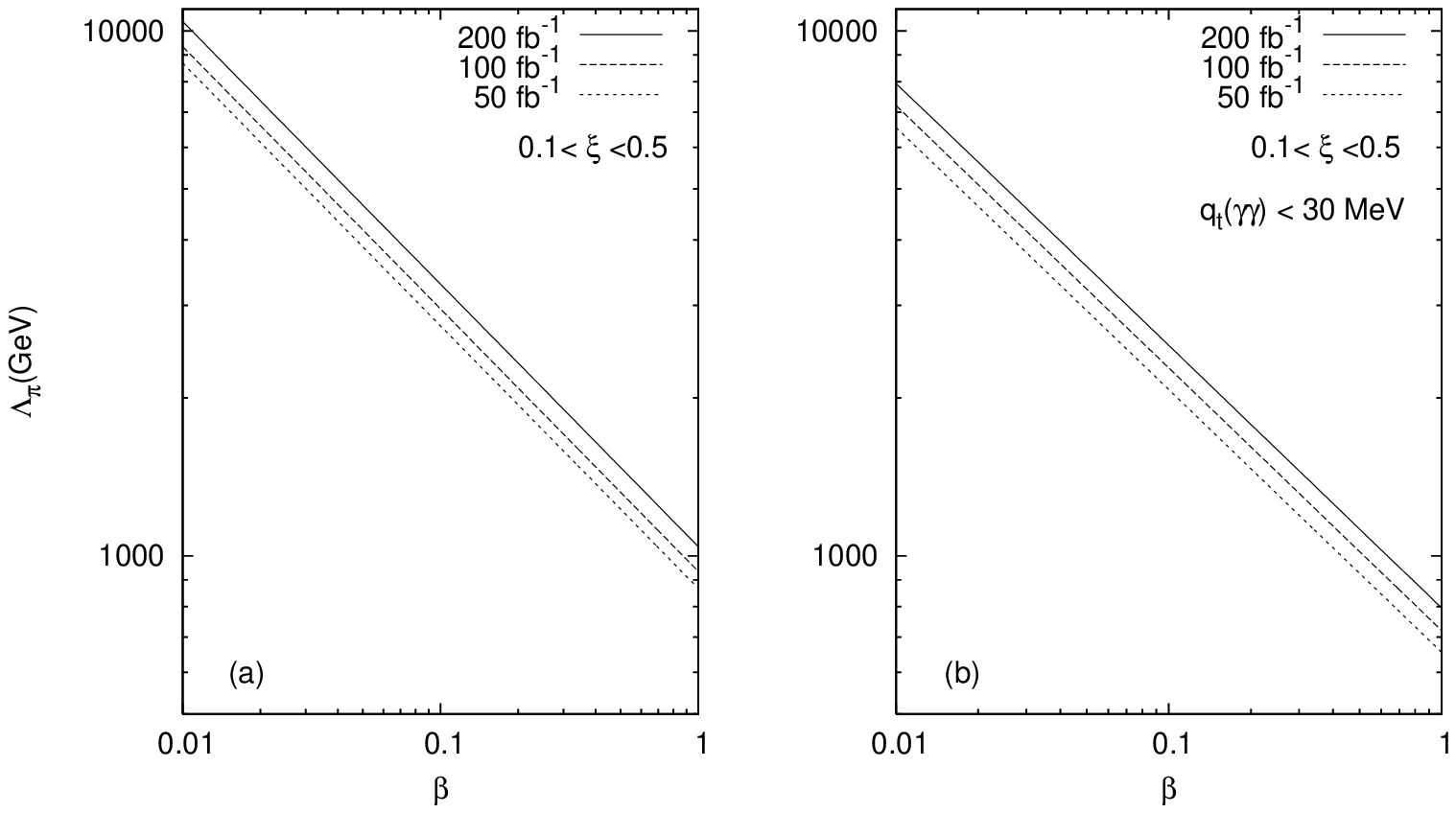}
\caption{95\% C.L. constraints  on the parameter plane
$\Lambda_{\pi}$ and $\beta$  at three different integrated LHC
luminosities; $50 fb^{-1}$, $100 fb^{-1}$ and $200 fb^{-1}$.
Curves in the  Fig.(b)  are obtained with
the cut on the transverse momentum of the
incoming photon pair. Excluded regions are defined
by the area below the curves.
\label{fig6}}
\end{figure}

Fig.\ref{fig6} shows 95\% C.L. search reaches
in the $\Lambda_{\pi} - \beta$ plane for an acceptance region of
$0.1 <\xi<0.5$ and LHC luminosities given before.
Calculations concerning the
exclusion contours for other acceptance limits
$0.0015<\xi<0.5$ and  $0.05<\xi<0.5$ were done.
As in the case of the ADD model these results show almost the
same curves with convenient transverse momentum cutoffs
on the individual final leptons. Thus we have not presented them
here.

\section{Conclusion}

Photon-photon collision at LHC with di-photon invariant mass $ W> 1$
TeV allows to study physics at TeV scale beyond the SM  with a
sufficient luminosity. There is no existing collider with this
property except the LHC itself. It is worth mentioning that two
lepton final state is the cleanest channel to search for any
deviation from the SM physics. For this reason, we have investigated
how the photon-induced two lepton final state can extend the bounds
on the model parameters of extra dimensions in the framework of the
ADD and RS models. Taking an acceptance region of $0.1 < \xi < 0.5$
we have obtained constraints on the fundamental scale $M_{D}$ in the
ADD model for a LHC luminosity interval 1-200 $fb^{-1}$. Exclusion
regions of the parameter pairs $\beta -m_{1}$ and
$\Lambda_{\pi}-\beta$  have been provided for the LHC luminosities
$50 fb^{-1}$, $100 fb^{-1}$ and $200 fb^{-1}$. Possible
contamination coming from baryon excitations can be removed by an
upper cut on the transverse  momentum of the incoming photon pair
$|\vec{q}_{1t}+\vec{q}_{2t}| <30$ MeV. An overall acceptance extends
between $\xi=0.0015$ and $\xi= 0.5$. We have showed that the region
$0.0015 < \xi < 0.5$ with a cut of $p_{t}>450-500$ GeV on final
leptons yields equivalent results to the region of $0.1 < \xi < 0.5$
with $p_{t}>10$ GeV. Excluded area of the model parameters that we
have found from the process $pp \to p\ell\ell p$ extends to wider
regions than the case of the colliders LEP and Tevatron
\cite{rizzo}. Certainly, the challenging process is the Drell-Yan
process at the LHC itself which will have the best sensitivity to
the KK tower parameters \cite{Dvergsnes:2004tw}.

\appendix
\section{}
Symbols $C^{\alpha\beta\rho\sigma}$ and
$D^{\alpha\beta\rho\sigma}$ that were used in the coupling
of KK states and photons in the text are defined as follows

\begin{eqnarray}
C^{\alpha\beta\rho\sigma}&&=\eta^{\alpha\rho}\eta^{\beta\sigma}
+\eta^{\alpha\sigma}\eta^{\beta\rho}-
\eta^{\alpha\beta}\eta^{\rho\sigma} \\
D^{\alpha\beta\rho\sigma}&&=\eta^{\alpha\beta}
k^{\sigma}_{1}k^{\rho}_{2}-(\eta^{\alpha\sigma}
k^{\beta}_{1}k^{\rho}_{2}+\eta^{\alpha\rho}
k^{\sigma}_{1}k^{\beta}_{2}-\eta^{\rho\sigma}
k^{\alpha}_{1}k^{\beta}_{2})-
(\eta^{\beta\sigma}
k^{\alpha}_{1}k^{\rho}_{2}+\eta^{\beta\rho}
k^{\sigma}_{1}k^{\alpha}_{2}-\eta^{\rho\sigma}
k^{\beta}_{1}k^{\alpha}_{2})
\end{eqnarray}

\end{document}